\documentclass[]{llncs}      % it must be
\usepackage[dvips]{graphicx} % if needed

\begin{document}             % it must be

\newcommand{\ga}{\mbox{ \raisebox{-.5ex}{$\stackrel{\textstyle >}{\sim}$} }}

\title{Statistical Properties of Dissipative MHD Accelerators}

\author{Kaspar Arzner\inst{1}, Loukas Vlahos\inst{2}, 
        Bernard Knaepen\inst{3} and Nicolas Denewet\inst{3}}

\institute{
        Paul Scherrer Institut \\
	Laboratory for Astrophysics \\
        CH-5232 Villigen PSI, Switzerland \\
        arzner@astro.phys.ethz.ch \\[2mm]
\and
        Aristotle University \\
        Institute of Astronomy, Dept. of Physics \\
        54006 Thessaloniki, Greece \\
	vlahos@astro.auth.gr \\[2mm]
\and
	Universit\'e Libre de Bruxelles \\
	Mathematical Physics Dept. \\
	CP231, Boulevard du Triomphe \\
	1050 Bruxelles, Belgium \\
	bknaepen@ulb.ac.be}

\maketitle
\index{Arzner, Kaspar}
\index{Vlahos, Loukas}
\index{Knaepen, Bernard}

\markboth{Arzner et al.}
{Particle Acceleration}

\begin{abstract}
We use exact orbit integration to investigate particle acceleration in a Gauss field proxy of 
magnetohydrodynamic (MHD) turbulence. Regions where the electric current exceeds a critical threshold are declared 
to be `dissipative' and endowed with super-Dreicer electric field ${\bf E}_\Omega = \eta {\bf j}$. In this
environment, test particles (electrons) are traced and their acceleration to relativistic 
energies is studied. As a main result we find that acceleration mostly takes place within the 
dissipation regions, and that the momentum increments have heavy (non-Gaussian) tails, while the waiting 
times between the dissipation regions are approximately exponentially distributed with intensity 
proportional to the particle velocity. No correlation between the momentum increment and the 
momentum itself is found. Our numerical results suggest an acceleration scenario with ballistic transport
between independent `black box' accelerators.
\end{abstract}

\section{Introduction}

Astrophysical high-energy particles manifest as cosmic rays or, indirectly, as
radio waves, X-rays, Gamma rays. These often occur in transients, and with distinctly
non-equilibrium energy distributions. A prominent source of sporadic radio- and X-ray emission is the Sun during the
active phase of its 11-year cycle. Among the numerous mechanisms proposed for accelerating solar particles to high energies 
(see \cite{miller97} for an overview), stochastic ones attracted particular 
attention because they require generic input data and do not rely on special geometrical 
assumptions. In stochastic acceleration \cite{karimabadi87,miller87,miller96,miller97b}, particles move in random electromagnetic fields, 
where they become repeatedly deflected and, on average, accelerated. The electromagnetic 
fields are thought to arise from magneto-hydrodynamic (MHD) turbulence (e.g., \cite{biskamp00}), 
perhaps excited by the broadband echo of a magnetic collapse. The turbulence may
host shocks and other forms of dissipation if critical velocities \cite{treumann97} or 
electric current densities \cite{papadopoulos79,parker93} are exceeded. Associated with dissipation are
(collisional or anomalous) resistivity and non-conservative electric fields, 
which sustain, locally, the electric current against dissipative drag in order to meet 
the global constraints. However, a detailed balance on the level of individual charge carriers
is impossible because the dissipative drag depends on particle position {\it and} velocity,
whereas the electric field is a function of position only. Thus the electric field 
may compensate the bulk drag, but a (high-energy) population can be left
over and exposed to acceleration \cite{dreicer60}. This lack of detailed 
balance is in the heart of dissipative acceleration mechanisms. 
In plasmas, dissipation occurs at `ruptures' of the magnetic structure, and 
is therefore localized around critical points of the magnetic field.

The above scenario, first envisaged by Parker \cite{parker83} for the solar atmosphere, has since been explored 
in a large number of numerical studies \cite{matthaeus86,ambrosiano88,vlahos97,biskamp00,dimitruk03,moriyashu04,arzner04}. 
On the theoretical side, most stochastic acceleration theories \cite{karimabadi87,karimabadi90,schlickeiser02} 
base on Fokker-Planck approaches, thus transferring two-point functions of
the electromagnetic fields into drift and diffusion coefficients of particles by
probing the fields along unperturbed trajectories. Dynamical particle averages are then 
replaced by field ensemble averages, neglecting the fact that real particles move in {\it one} 
realization of the random field. As a result, diffusive behaviour may be predicted even if 
particles are trapped in a single realization of the random field. 

In order to investigate the full diversity of orbit behaviour one must resort to numerical simulations.
In the present contribution we analyze the behaviour of test particles in resistive MHD turbulence
with localized dissipation regions, with particular emphasis on the validity of a Fokker-Planck description \cite{gardiner85}. 
We use here exact orbit integration, and thus avoid any guiding centre approximations \cite{littlejohn82,littlejohn83}.
The price for rigorosity is computational cost, which makes the scheme only feasible with the aid of 
high-performance computing.

\section{Acceleration Environment}

The MHD turbulence has been been modeled by full 3D spectral MHD 
simulations and by Gauss field proxies \cite{adler81,arzner04}. We concentrate here on
the latter, which is computed from the vector potential ${\bf A}({\bf x},t)$ = 
$\sum_{\bf k} {\bf a}_{\bf k} \cos ({\bf k} \cdot {\bf x} -\omega({\bf k})t-\phi_{\bf k})$ by means of 
tabulated trigonometric calls. This allows to continuously determine the fields at the exact particle
position, and avoids any real-space discretization artifacts, but
the computational overhead restricts the ${\bf k}$ sum to a few 100 Fourier 
modes ${\bf a_k}$. They are taken from the shell ${\rm min}(l_i^{-1})$$<$$|{\bf k}|$$<$ 
$10^{-2} r_L^{-1}$ with $r_L$ the rms thermal ion Larmor radius and $l_i$ the outer scale of 
the power spectral density $\langle |{\bf a}_{\bf k}|^2 \rangle \propto \, (1 + l_x^2k_x^2+l_y^2k_y^2+l_z^2k_z^2)^{-\nu}$.
The electromagnetic fields are then obtained from

\begin{eqnarray} 
{\bf B} & = & \nabla \times {\bf A} \label{B} \\ 
{\bf E} & = & - \partial_t {\bf A} + \eta ({\bf j}) \, {\bf j} \, , \label{E} 
\end{eqnarray} 

where $\mu_0 {\bf j}$ = $\nabla \times {\bf B}$ and $\eta({\bf j})$ = $\eta_0 \, \theta(|{\bf j}|-j_c)$ 
is an anomalous resistivity switched on above the critical current $j_c \sim e n c_s$. Here, $c_s$ and $n$ 
are the sound speed and number density of the background plasma. The Gauss field ${\bf A}$ must satisfy 
the MHD constraints 

\begin{equation} 
{\bf E} \cdot {\bf B} = 0 \;\;\; \mbox{if} \;\;\; \eta({\bf j}) = 0 
\;\;\;\;\; \mbox{and} \;\;\;\; E/B  \sim v_A \label{MHD} 
\end{equation} 

with $v_A$ the Alfv\'en velocity. Equation (\ref{MHD}) can be achieved in several ways. For instance,
one can use Euler potentials of which only one is time-dependent, or force ${\bf A}$ to
point along a single direction. A somewhat more flexible way, used here, is
axial gauge ${\bf a}_{\bf k} \cdot {\bf v}_A$ = 0 with dispersion relation $\omega({\bf k})$ = 
${\bf k} \cdot {\bf v}_A$. A constant magnetic field $B_0$ along ${\bf v}_A$ can be included 
without violating Eq. (\ref{MHD}), and we set $|{\bf v}_A|^2$ = $(B_0^2+\sigma_B^2)/(\mu_0 n m_p)$
with $\sigma_B$ the rms magnetic fluctuations and $m_p$ the proton rest mass. In the present simulation,
${\bf v}_A$ and the background magnetic field are along the $z$ direction. The total magnetic field 
${\cal B}$ = $\sqrt{B_0^2+\sigma_B^2}$ is a free parameter, which defines the scales of the 
particle orbits. In order to represent coronal turbulence we choose $v_A \sim 2 \cdot 10^6$ m/s, $\nu$ = 1.5, 
${\cal B} \sim 10^{-2}$T, $\sigma_B$ = $10^{-2}$T, $B_0$ = 10$^{-3}$T, and $l_x$ = $l_y$ = 10$^3$m, $l_z$ = 
10$^4$m. The current threshold $j_c$ is exceeded in about 7\% of the total volume. Note that our choice
represents strong ($\sigma_B \gg B_0$) and anisotropic ($l_z \gg l_x, l_y$) turbulence. To embed our
simulation in the real solar atmosphere one should associate $l_z$ with the radial direction in order
to reproduce the predominant orientation of coronal filaments.

\begin{figure}[h]
\begin{center}
\includegraphics[height=9.5cm,width=13cm]{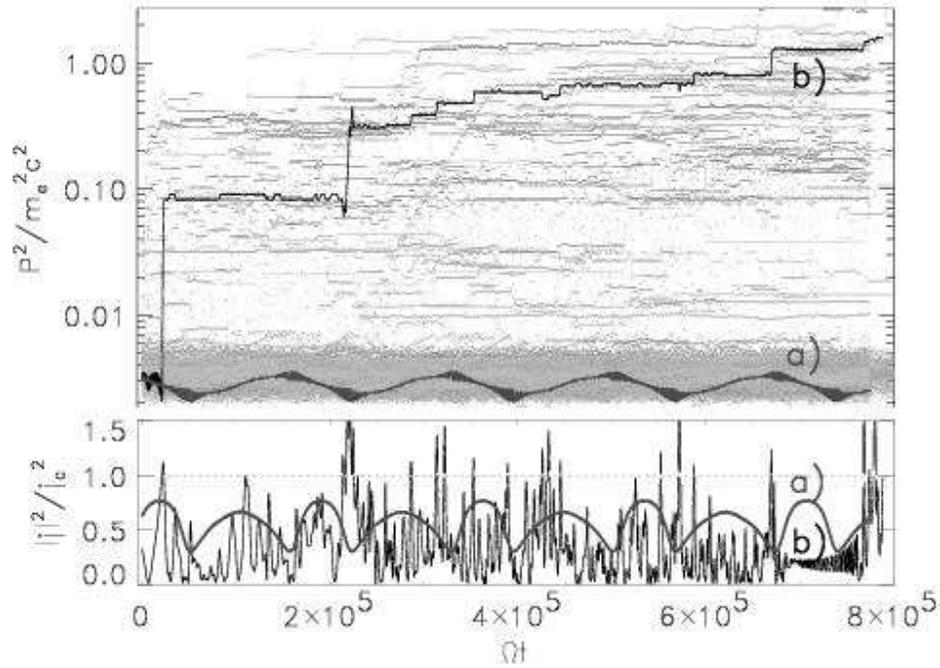}
\end{center}
\vspace{-7mm}
\caption{
Evolution of electron kinetic momentum. Top panel: 200 sample orbits; adiabatic (a), and
accelerated (b) cases. Bottom panel: electric current density along the orbits a) and b).
The critical current density ($|{\bf j}|>j_c$) is marked by dotted line. The present 
simulation is an extension of the simulation of \cite{arzner04}.
\label{electrons_fig}}
% this figure was created by /afs/psi.ch/user/a/arzner/loukas/lifecycle/traj_f90/merlin/idl/plot_electrons_para04.pro
\end{figure}

\begin{figure}[h]
\hspace{-1cm}
\includegraphics[height=9cm,width=13cm]{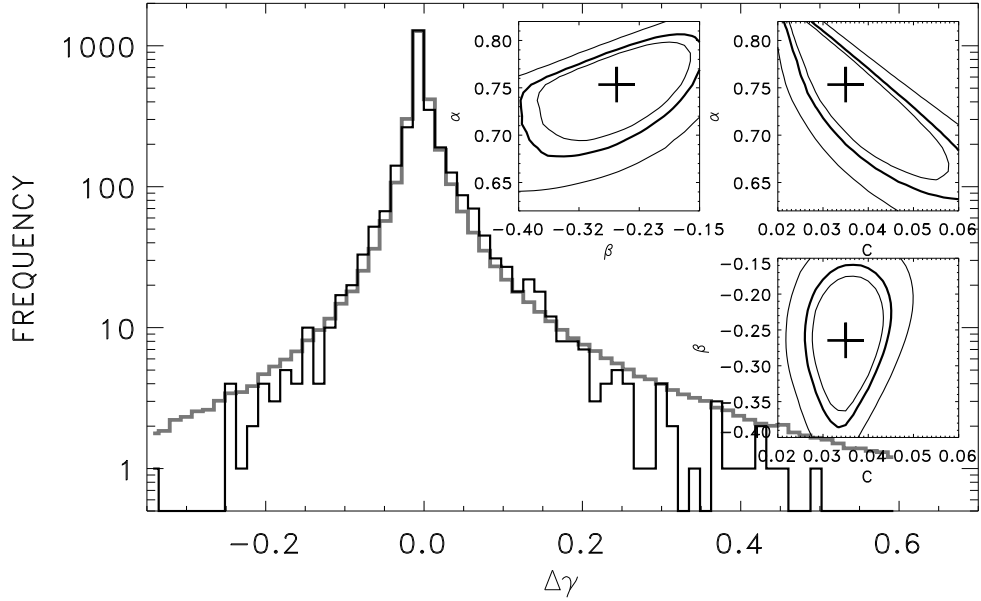}
\caption{Frequency distribution of the energy jumps $\Delta \gamma$ of Fig. \protect\ref{electrons_fig} 
(black line), together with a best-fit L\'evy density (gray line) with parameters 
$\alpha_0$ = 0.75, $\beta_0$ = $-$0.26, and $C_0$ = 0.035 (crosses). Inlets: cuts of the likelihood surface at 
($\alpha_0,\beta_0,C_0$). The 99\% confidence level is marked boldface.
\label{de_fig}}
% this figure was created by /afs/psi.ch/user/a/arzner/loukas/lifecycle/traj_f90/merlin/idl/de_levy.pro
\end{figure}

\begin{figure}[h]
\hspace{-1cm}
\centerline{
\includegraphics[height=6cm,width=7cm]{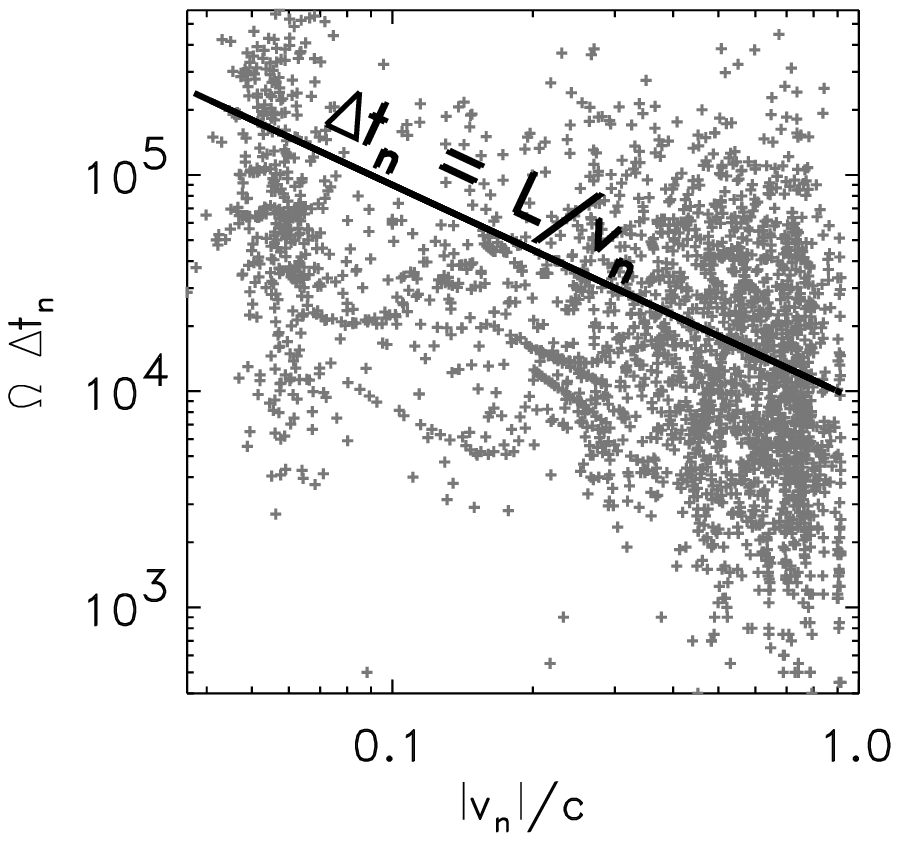}
\includegraphics[height=6cm,width=7cm]{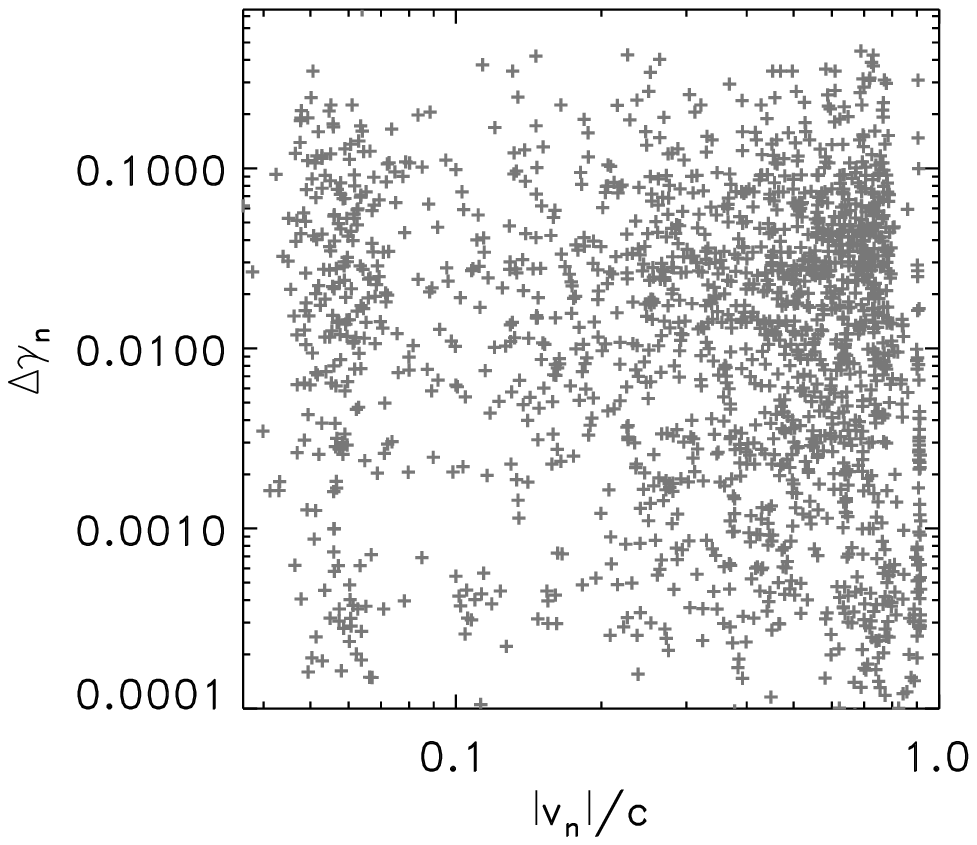}}
\caption{Left: travel times $\Delta t_n = t_{n+1}-t_n$ between acceleration regions versus velocity $v_n$. 
Right: energy gain $\Delta \gamma_n = \gamma_{n+1}-\gamma_n$ versus velocity. While $\Delta t_n$ scales with
inverse velocity (solid line: best-fit), there is no clear trend in the energy gain $\Delta \gamma_n$.}
\label{dt_dE_v_fig}
% this figure was created by /afs/psi.ch/user/a/arzner/loukas/lifecycle/traj_f90/merlin/idl/dt.pro
\end{figure}

\begin{figure}[h]
\hspace{-1cm}
\includegraphics[height=7cm,width=10.5cm]{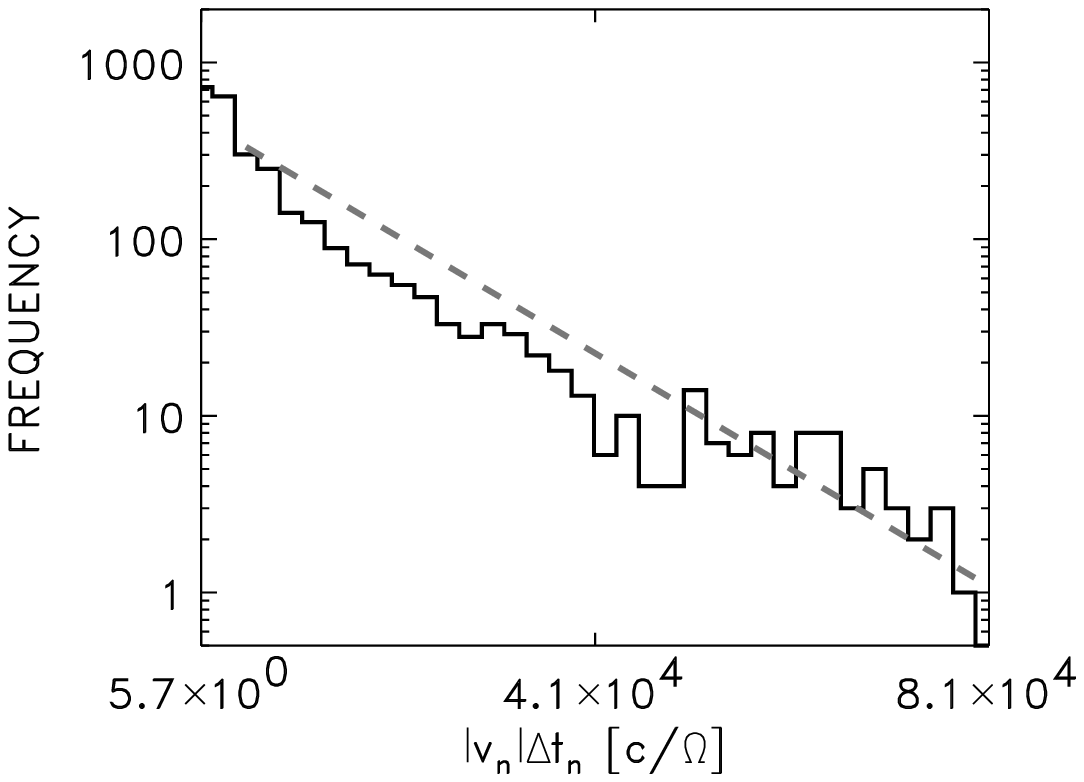}
\caption{Histogram of the quantity $|v_n| \Delta t_n$, together with an exponential fit (dashed).}
\label{dt_hist_fig}
% this figure was created by /afs/psi.ch/user/a/arzner/loukas/lifecycle/traj_f90/merlin/idl/dt.pro
\end{figure}

\section{Particle Dynamics} 

\subsection{Physical Scaling}

Time is measured in units of the (non-relativistic) gyro period $\Omega^{-1}$=$m/q{\cal B}$; velocity in units 
of the speed of light; distance in units of 
$c\Omega^{-1}$. Particle momentum is measured in units of $mc$; the vector potential in units of $mc/q$; the 
magnetic field in units of ${\cal B}$; the electric current density in units of 
$\Omega {\cal B} / (\mu_0 c)$, so that the dimensionless threshold current is 
$j_c' = (m/m_p) c_s c / v_A^2$; and the electric field is measured in units of $c {\cal B}$, so 
that the dimensionless Dreicer \cite{dreicer60} field is $E_D'$ = $(v_e'/\tau') (m_e/m)$ 
with $v_e'$ the electron thermal velocity and $\tau'$ the 
electron-ion collision time. The dimensionless equations of motion are 

\begin{eqnarray} 
\frac{d {\bf x}'}{d t'} & = & {\bf v}' \label{dxdt} \\ 
\frac{d(\gamma {\bf v'})}{dt'} & = & {\bf v}' \times {\bf B}' - \frac{\partial {\bf A}'}{\partial t} 
	+  \eta'(|{\bf j}'|) \, {\bf j}'
\label{dPdt} 
\end{eqnarray} 

with $\gamma$ the Lorentz factor, ${\bf B}' = \nabla' \times {\bf A}'$, and 
${\bf j}' = \nabla' \times {\bf B}'$ the electric current. 
The dimensionless resistivity $\eta'$ is characterized by the resulting dissipative 
electric field ${\bf E}_\Omega = \eta_0 {\bf j}$ relative to the Dreicer field $E_D$.
We chose $\eta'$ such that $E_\Omega/E_D$ $\sim$ $10^4$.

\subsection{Particle Initial Conditions}

We consider electrons as test particles. The initial positions are uniformly distributed in space, and the
velocities are from the tail $v$ $\ge$ 3 $v_{th}$ of a maxwellian of 10$^6$ K, which is typical
for the solar corona. Coulomb collisions are neglected, which is a good approximation once the 
acceleration has set on, but is not strictly correct in the beginning of the simulation.

\subsection{Numerical Implementation and Simulation Management}

Equations (\ref{dxdt}) and (\ref{dPdt}) are integrated by traditional leapfrog and Runge-Kutta 
schemes. The test particle code is written in FORTRAN 90/95 and compiled by the Portland Group's
Fortran 90 compiler ({\tt pgf90}). Diagnostics and visualization uses IDL
as a graphical back-end. The code is run on the MERLIN cluster of the Paul Scherrer Institut, and
on the ANIC-2 cluster of the Universit\'e libre de Bruxelles. The ANIC-2 cluster has 
32 single Pentium IV nodes, a total of 48 Gbyte memory, and Ethernet connections. The MERLIN cluster consist 
of 56 mostly dual Athlon nodes with a total of 80 GByte memory, operated
under Linux and connected by Myrinet and Ethernet links. MERLIN jobs are managed by the Load 
Sharing Facility (LSF) queueing system. Parallelization is done on a low level only, 
with different (and independent) test particles assigned to different CPU's. MPICH/MPI is
used to ensure crosstalk-free file I/O. The field data are computed on each CPU for the
actual particle position. Random numbers are needed in the generation of the Fourier amplitudes and
-phases of the electromagnetic fields, and in the particle initial data; they are taken from the intrinsic 
random number generator of {\tt pgf90}.

\section{Diagnostics and Results}

In order to characterize the relativistic acceleration process we consider the evolution of the
kinetic momentum ${\bf P}' = \gamma {\bf v}'$. This quantity is directly incremented by the equation of motion 
(\ref{dPdt}), and -- ignoring quantum effects -- can grow to arbitrarily large values, so that it can serve as 
a diagnostics of diffusive behaviour. Alternatively we may use the kinetic energy $\gamma$.

The results of the orbit simulations are shown in Figs. \ref{electrons_fig} - \ref{dt_hist_fig}. 
When initially super-thermal ($v \ga 3 v_{th}$) electrons move in the turbulent electromagnetic fields 
(Eqns. \ref{B}, \ref{E}), some of them may become stochastically accelerated. From a population
of 600 electrons we find that 35\% of the particles are accelerated, while the other 65\% remain adiabatic 
\cite{littlejohn83,buechner89} during the simulation ($\Omega t \le 8 \cdot 10^5$). The two cases are 
illustrated in Fig. \ref{electrons_fig} (top). The orbit a) conserves energy adiabatically during the
whole simulation, while the orbit b) does not. The orbits of 200 randomly chosen particles are also shown 
(gray) to trace out the full population. The bottom panel of Fig. \ref{electrons_fig} shows the 
the electric current density along the orbits a) and b). Time intervals where the critical current (doted line)
is exceeded correspond to visits to the dissipation regions. As can be seen, acceleration (or deceleration) occurs 
predominantly within the dissipation regions. Accordingly, the orbit b) which never enters a dissipation region remains
adiabatic. As a benchmark we have set $\eta_0$ = 0 and found that no acceleration takes place at all,
thus reproducing the `injection problem' \cite{cargill01}. Smaller $\eta'$ yield smaller (than 35\%)
fractions of accelerated particles.

A glance at graph a) of Fig. \ref{electrons_fig} shows that ${\bf P}'(t')$ is poorly represented by a 
Brownian motion \cite{gardiner85} with continuous sample paths. Rather, ${\bf P}'$ changes intermittently and
in large jumps. Indeed, if we consider the energy change $\Delta \gamma_n = \gamma_{n+1}-\gamma_n$ across a 
dissipation region, we find that its distribution  $P(\Delta \gamma_n)$ has heavy tails and a convex shape which deviates 
from a Gaussian law (Fig. \ref{de_fig} black line). In order to characterize $P(\Delta \gamma_n)$ we have 
tried to fit it by a (skew) L\'evy stable distribution $P_L(x)$. The latter is defined in terms of its Fourier 
transform \cite{lukacs60}

\begin{equation}
\phi_L(s) = \exp \Big\{ - C |s|^\alpha \Big(1 + i \beta \frac{s}{|s|} \tan \frac{\pi \alpha}{2} \Big) \Big\}
\label{phi_Levy}
\end{equation}

with $0 < \alpha \le 2$, $-1 \le \beta \le 1$, and $C > 0$. The parameter $\alpha$ determines the asymptotic decay 
of $P_L(x) \sim x^{-1-\alpha}$ at $x \gg C^{1/\alpha}$, and $\beta$ determines its skewness. The probability density
function (PDF) belonging to Eq. 
(\ref{phi_Levy}) has the `stability' property that the sum of independent identically L\'evy distributed variates is 
L\'evy distributed as well. While rapidly converging series expansions \cite{lukacs60} of $P_L(x)$ are 
available for $1 < \alpha \le 2$ or large arguments $x$, the evaluation of $P_L(x)$ at small arguments and $\alpha < 1$
is more involved. We use here a strategy where $P_L(x)$ is obtained 
from direct computation of the Fourier inverse $P_L(x) = (2\pi)^{-1} \hspace{-1mm} \int e^{-isx} \phi_L(s) \, ds$, 
with the integrand split into regimes of different approximations. At small $|s|$, both the exponential in
$\phi_L(s)$ (Eq. \ref{phi_Levy}) and the Fourier factor $e^{-isx}$ are expanded; at larger 
$|s|$, $\phi_L(s)$ is piecewise expanded while $e^{-isx}$ is retained. In both cases, the
$s$-integration can be done analytically, and the pieces are summed numerically. The resulting (Poisson) 
maximum-likelihood estimates of the parameters $(\alpha, \beta, C)$ are $\alpha_0$ = 0.75, $\beta_0$ = $-$0.26, 
$C_0$ = 0.035. The predicted frequencies are shown in Fig. \ref{de_fig} (gray line), and inlets represent 
sections of constant likelihood in ($\alpha,\beta,C$)-space, with the 99\% confidence region enclosed by boldface 
line. The finding $\alpha_0 < 2$ agrees with the presence of large momentum jumps.

In a next step we have investigated the waiting times $\Delta t_n = t_{n+1}-t_n$ between subsequent 
encounters with the dissipation regions. Simple ballistic transport between randomly positioned dissipation
regions would predict a PDF of the form $f(|v_n| \Delta t_n)$ with $v_n$ the particle velocity and
$f(x)$ the PDF of distances between (magnetically connected) dissipation regions. (There is no Jacobian
$d (\Delta t) / d x$ since we are dealing with discrete events.)
This is in fact the case. Figure \ref{dt_dE_v_fig} (left) shows a scatter plot of the actual velocity $v_n$ versus 
waiting time $\Delta t_n$. Gray crosses represent all simulated encounters with dissipation regions,
including all particles and all simulated times. There is a clear trend for $\Delta t_n$ to scale with $v_n^{-1}$, and the 
black solid line represents a best fit of the form $\Delta t_n = L / v_n$ with $L = 9 \cdot 20^3 \, c/\Omega$.
When a similar scatter plot of velocity versus energy gain is created (Fig. \ref{dt_dE_v_fig} right), then
no clear correlation is seen: the energy gain is apparently independent of energy. In this sense the dissipation 
regions `erase' the memory of the incoming particles. Returning to the
waiting times, we may ask for the shape of the function $f(|v_n| \Delta t_n)$. This can be
determined from a histogram of the quantity $|v_n| \Delta t_n$ (Fig. \ref{dt_hist_fig}, solid line). The decay is
roughly exponential (dashed: best-fit), although the limited statistics does certainly not allow to
exclude other forms.

\section{Summary and Discussion}
 
We have performed exact orbit integrations of electrons in a Gauss field proxy of MHD turbulence 
with super-Dreicer electric fields localized in dissipation regions.
It was found that the electrons remain adiabatic (during the duration of the simulation) if no 
dissipation regions are encountered, and can become accelerated if such are met. The resulting
acceleration is intermittent and is not well described by a diffusion process, 
even if the underlying electromagnetic fields are Gaussian. On time scales which are
large compared to the gyro time, the kinetic momentum performs a L\'evy flight rather 
than a classical Brownian motion. The net momentum increments in the dissipation regions are independent
of the in-going momentum, and have heavy tails which may be approximated by a stable law of index 0.75.
The waiting times between subsequent encounters with dissipation regions are approximately exponentially
distributed, $P(\Delta t) \sim e^{- v \Delta t / L}$, indicating that the dissipation regions
are randomly placed along the magnetic field lines. This one-dimensional Poisson behaviour is most 
likely caused by the Gauss field approximation, and the waiting times in true MHD turbulence are
expected to behave differently. An ongoing study is devoted to these questions, and results 
will be reported elsewhere. In summary, our numerical results suggest that the acceleration process 
may be modeled by a continuous-time random walk with finite or infinite mean waiting 
time, and infinite variance of the momentum increments. Such models can be described in
terms of fractional versions \cite{uchaikin00,meerschaert02} of the Fokker-Planck equation.

%%%%%%%%%%%%%%%%% REFERENCES %%%%%%%%%%%%%%%%%%%%%

\end{document}